\documentclass[useAMS,usenatbib]{mn2e} 
\usepackage{epsfig}

\def\spose#1{\hbox to 0pt{#1\hss}} 
\def\lta{\mathrel{\spose{\lower 3pt\hbox{$\mathchar"218$}}      
     \raise 2.0pt\hbox{$\mathchar"13C$}}}      
\def\gta{\mathrel{\spose{\lower 3pt\hbox{$\mathchar"218$}}      
     \raise 2.0pt\hbox{$\mathchar"13E$}}}      
   
\title[The Chemical Evolution of Dwarf Spheroidals]{The 
Chemical Evolution of Dwarf Spheroidal Galaxies: Dissecting the 
Inner Regions and their Stellar Populations}
   
\author[Marcolini et al.] 
{A. Marcolini$^{1}$, A. D'Ercole$^{2}$, G. Battaglia$^{3,4}$ and
       B.K. Gibson$^{1,5}$ \\  
       $^1$ Centre for Astrophysics, University of Central Lancashire, 
       Preston, Lancashire, PR1 2HE, United Kingdom \\
       $^2$ Osservatorio Astronomico di Bologna,   
       via Ranzani 1, 40127 Bologna, Italy \\
       $^3$ European Organization for Astronomical Research in the Southern 
       Hemisphere, K. Schwarzschild-Str. 2, 85748 Garching, Germany \\
       $^4$ Kapteyn Institute, University of Groningen, Postbus 800, 
       9700AV Groningen, the Netherlands \\
       $^5$ School of Physics, University of Sydney,
       NSW, 2006, Australia}

\date{Accepted ..., Received ...; in original ...}     
   
\pagerange{\pageref{firstpage}--\pageref{lastpage}}     
\pubyear{2008}     
     
\begin{document}     
  
\maketitle     
     
\label{firstpage}     
     
\begin{abstract}    

Using 3-dimensional hydrodynamical simulations of isolated 
dwarf spheroidal galaxies (dSphs), we undertake an analysis of the
chemical properties of their inner regions, identifying the 
respective roles played by Type~Ia (SNe~Ia) and Type~II (SNe~II)
supernovae.  The effect of inhomogeneous pollution from SNe~Ia is
shown to be prominent within two core radii, with the stars forming
therein amounting to $\sim$20\% of the total.
These stars are relatively iron-rich and $\alpha$-element-depleted 
compared to the stars forming in the rest of the galaxy. 
At odds with the projected stellar velocity
dispersion radial profile, the actual 3-dimensional one shows a depression 
in the central region, where the most metal-rich (ie. [Fe/H]-rich) 
stars are partly segregated. 
This naturally results in two different stellar populations,
with an anti-correlation between [Fe/H] and velocity dispersion, in the
same sense as that observed in the Sculptor and Fornax dSphs.
Because the most iron-rich stars in our model are also the most
$\alpha$-depleted, a natural prediction and test of our model is that the 
same radial segregation effects should exist between [$\alpha$/Fe] and
velocity dispersion.

\end{abstract}    
 
    
\section{Introduction} 
\label{sec:introduction} 

Hierarchical structure formation models predict that massive galaxies
formed through the continuous accretion of numerous satellites, and
that such a process, at an admittedly lower rate, should be an ongoing
one.  Simulations of Milky Way (MW)-like galaxies show stellar halos
consisting predominantly of stellar debris from disrupted satellites
that is kinematically and spatially distinct from the population of
surviving satellites \citep[e.g.][]{bullock2001}.  This is because the
survival of a satellite is significantly biased toward outlying
lower-mass systems with less eccentric orbits that have fallen into
the galaxy at later times \citep[e.g][]{moore2006,sales2007}. It is
not clear if the star formation histories (SFH) and the gas contents
of these systems are regulated by internal feedback
\citep[e.g.][]{dekel1986} and/or (more likely) by the tidal
interaction with the MW and by the ram pressure stripping due to the
hot halo of the Galaxy \citep[e.g.][]{mayer2006}.

If these satellites, or dwarf galaxies, contributed to the build-up of
the MW, it becomes fundamentally important to reconstruct how the
heavy elements have been produced over time in {\it these} systems, in
order to understand the evolution of our own Galaxy. With
high-resolution spectrographs on 8m-class telescopes, it is now
possible to take detailed spectra of individual stars in nearby dwarf
galaxies, thereby shedding light on the latters' chemical enrichment
and star formation histories, and consequently, the detailed
accounting of the number and and type of supernovae that occurred in
the past \citep[e.g.][]{matteucci1985}. For this reason, a rich
literature exists devoted to studying the chemical evolution of the
Galactic satellites. In particular, several authors
\citep[e.g.][]{gilmore1991, carigi2002, ikuta2002, lanfranchi2004,
lanfranchi2007, robertson2005, fenner2006, gibson2007, carigi2008}
have made chemical models of dwarf spheroidals, employing primarily
the assumption of a single-zone framework - i.e., the interstellar
medium (ISM) and stellar distributions are assumed to remain
homogeneous in space throughout the calculation. As a consequence,
spatial inhomogeneities cannot be taken into account by these models,
and the values of the inferred chemical and hydrodynamical
characteristics can only be interpreted as spatial averages.

Three-dimensional hydrodynamical simulations of the feedback between
ISM and SNe within dSph-sized objects have been carried out by
\citet{mori2002} and by \citet{marcolini2006}. \citet{mori2002}
considered pre-galactic objects which later merge into galactic halos;
in this context, SNe~II explode within a few tens of millions of
years, their energy efficiently coupled to the ISM in such a manner as
to drive an outflow containing a large fraction of the ISM and of the
freshly produced metals. Moreover, the rapid timescale over which this
outflow is effected ensures a negligible contribution from
SNe~Ia.\footnote{Indeed, SNe~Ia are explicitly neglected within the
simulations of \citet{mori2002}.}  On the contrary, the SFHs inferred
for surviving dSphs are consistent with being less intense and more
prolonged (spanning several Gyrs, in some cases), suggesting that
these objects have been able to retain their gas for long periods.

\begin{figure*}    
\begin{center}    
\psfig{figure=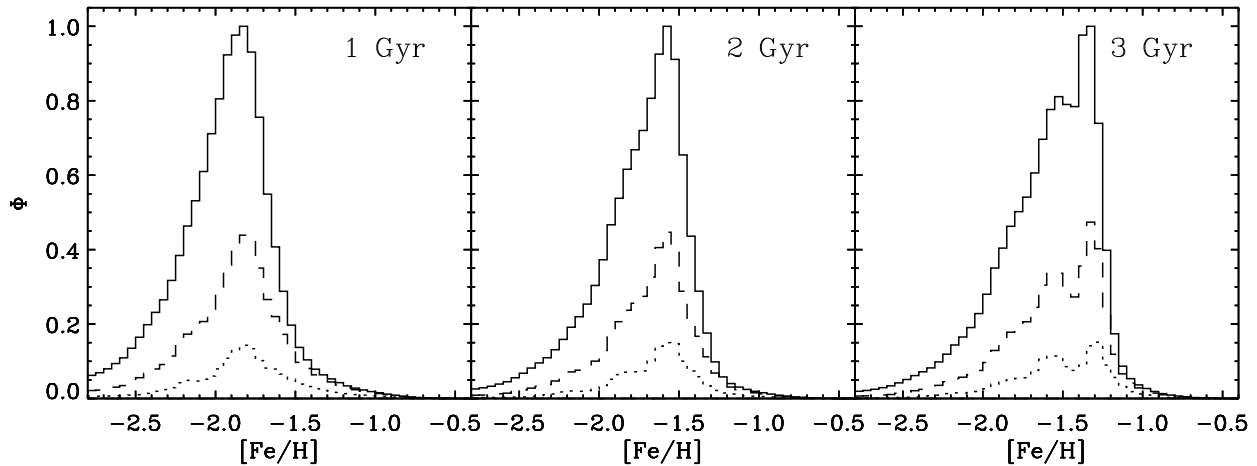}
\end{center}   
\caption{[Fe/H] distribution function of the long-lived stars for the
``entire dSph volume'' (solid line; $r_{\star, \rm c}=130$ and
$r_{\star, \rm t}=650$ pc), and within two (dashed line) and one core
radii (dotted line), at three different times: 1, 2, and 3 Gyrs,
respectively.}
\label{fig:zstelle} 
\end{figure*}

\begin{figure*}    
\begin{center}    
\psfig{figure=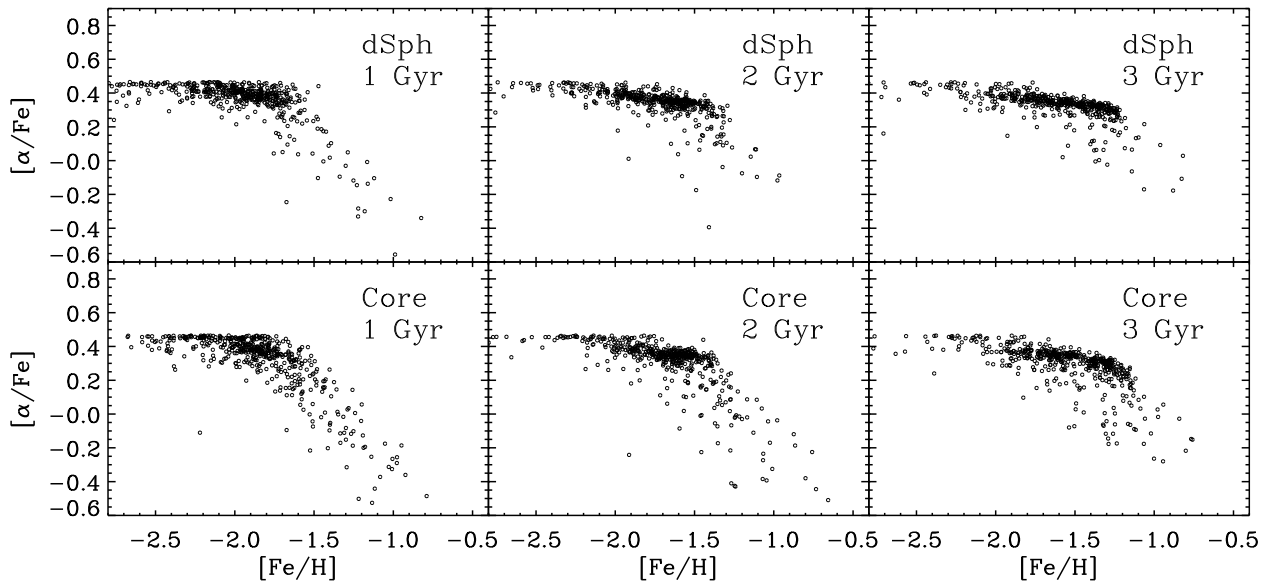,width=0.95\textwidth}    
\end{center}   
\caption{Upper panels: abundance ratio [$\alpha$/Fe] versus [Fe/H]
constructed for a random sample of $N_{\rm S}=500$ stars at three
different times, for the entire dSph simulated volume. Lower panels:
as above, but for stars forming within one core radius. Details on the
sampling method are given in footnote 2.}
\label{fig:zstelleoss} 
\end{figure*}

The simulations of \citet{marcolini2006} accommodated these prolonged
SFHs, allowing for the combined effects of both SNe~II and SNe~Ia.
The reason for such an extended SF is due to the effectiveness of
radiative losses which contrast the rapid expulsion of gas by SNe II,
and leaves it available for SF. As nowadays dSphs have no gas, this
scenario requires that the ISM is eventually lost by the interaction
with the Galaxy, probably through the combined effects of tidal forces
and ram-pressure stripping \citep[e.g][]{marcolini2003,
mayer2006}. Support for this picture is given by the fact that
isolated low mass dSphs such as Phoenix \citep{young2007} and Leo T
\citep{dejong2008} were able to form stars up to 100 Myr ago. A
further hint in the same direction is given by \citet{haines2007} who,
analysing a sample of $\sim$30,000 galaxies from the Sloan Digital Sky
Survey, found that the SFHs of dwarf galaxies are strongly dependent
on their local environment, the fraction of passively evolving
galaxies dropping from $\sim$70 per cent in dense environments, to
zero in the rarefied field.

While the model by \citet{marcolini2006} represents a noticeable
improvement from a hydrodynamical perspective, when compared to
traditional one-zone semi-numerical chemical evolution models, it
admittedly considered only the elements produced by supernovae. This
is not a serious drawback, in the sense that the elements synthesized
and distributed by SNe represent a significant fraction of all the
elements present today (modulo the contribution from low- and
intermediate-mass single stars), and the results of the model can be
readily tested against many observational datasets. It has been shown
that this model is consistent with many properties of the Draco dwarf
\citep{marcolini2006} and, with minimal assumptions, the chemical
properties of the globular cluster $\omega$-Centauri
\citep{marcolini2007}, which is believed to be the remnant of an
ancient dSph.

In the present paper, we analyze the results of the model by
\citet{marcolini2006} focusing on the chemical evolution of the stars
located in the central region of dSphs. It has been shown that in
several Local Group dSphs, including Sculptor and Fornax, the chemical
properties of the stars forming within the galactic core can differ
substantially from those situated at larger radii
\citep{tolstoy2003,tolstoy2004, battaglia2006}.  In Section
\ref{sec:genpi}, we summarise the characteristics of the
\citet{marcolini2006} simulation. We then discuss our results in
Section \ref{sec:res}, compare them to observational datasets in
Section \ref{sec:data}, and draw our conclusions in Section
\ref{sec:discussion}.

\section{General Picture}
\label{sec:genpi}

\citet{marcolini2006} carried out three-dimensional hydrodynamical
simulations of isolated dSphs, emphasising the different role played
by SNe~Ia and SNe~II in the chemical and dynamical evolution of the
system, in the presence of a prolonged intermittent SFH.  Although the
model has been employed to study the general characteristics of dSphs,
it was built initially to explore the origin and evolution of
the Draco dSph.  In this Section we
summarise the main characteristics of the model and its general
behaviour. Details of its rationale, as well as the choice of specific
parameters, can be found in \citet{marcolini2006}.

\subsection{The galaxy model}  
\label{sec:model} 

At first stars are absent, and the galactic potential is given by a
static dark halo whose density radial profile (truncated at $r_{\rm
h,t}$) is given by:

\begin{equation}   
\rho_{\rm h}(r)=\rho_{\rm h,0} \left [1+\left (\frac{r}{r _{\rm h,c}}\right )^2\right ]^{-1}.   
\label{equ:halo_dark}    
\end{equation}   

\noindent
The values of the halo parameters are given in Table \ref{tab:galaxy},
and lead to a halo mass $M_{\rm h}=6.2\times 10^7$ $M_{\odot}$.

Initially we place the ISM in isothermal equilibrium within the
potential well, with a temperature of $T_{\rm ISM} \sim T_{\rm
vir}=10^4$ K. The initial gas mass is $M_{\rm ISM} = 0.18 M_{\rm h}$,
according to the baryonic fraction given by \citet{spergel2007}, which
amounts to 1.1 $\times 10^7$ M$_{\odot}$. The contribution of this
baryonic mass to the gravity is neglected.

As the stars form (see next Section), they are randomly distributed in
space following a King profile (truncated at $r_{\rm \star,t}$):

\begin{equation}   
\rho_{\star}(r)=\rho_{\star, \rm 0} \left [1+ \left (\frac{r} 
{r _{\star, \rm c}}\right)^2 \right ]^{-3/2}. 
\label{equ:star_density}    
\end{equation}   
\noindent
After 3 Gyr (the duration of our SFH), a total stellar mass
$M_{\star}=5.6\times 10^5$ M$_{\odot}$ is formed. Assuming a stellar
mass to light ratio in the V band $M_{\star}/L_{\rm V}=2 \rm
M_{\odot}/L_{\odot}$, the mass to light ratio inside the stellar tidal
radius is $\rm M_{\odot}/L_{\rm V,\odot}=80$, in good agreement with
observations \citep[e.g.][]{mateo1998}.

The complete list of parameters concerning the galactic model is given
in Table~\ref{tab:galaxy}. We refer the reader to \citet{marcolini2006} 
for a detailed description of the model.

\subsection{SFH and SN explosions}
\label{sec:snmodel}

The SFH in the model is given by a sequence of 50 istantaneous bursts
of identical intensity separated by quiescent periods of 60 Myr. The
forming stars are randomly located within the galaxy following the
distribution given by Eq. \ref{equ:star_density}.

The SNe II associated to the freshly formed stars explode at a
constant rate for a period of 30 Myr (the lifetime of a 8 M$_{\odot}$
star, the least massive SN II progenitor) after each stellar burst.

The SN Ia rate for a single population decreases in time after an
initial rise.  We adopted the time-dependent rate given by
\citet{matteucci2001}, according to the Single-Degenerate scenario.
   
Each SN explosion is stochastically placed into the galaxy according
to its radial probability proportional to the stellar mass within that
radius $r$: $ P(r)=M_{\star}(r)/M_{\star}$.

Finally, we assume that each SN II ejects a mean mass $M_{\rm SN II,
ej}=10 \; \rm M_{\odot}$, and each SN Ia ejects $M_{\rm SNIa, ej}=1.4
\; \rm M_{\odot}$.  Every SN II expels 1.0 M$_{\odot}$ of oxygen and
$0.07$ M$_{\odot}$ of iron while each SN Ia ejects 0.15 M$_{\odot}$ of
oxygen and 0.74 M$_{\odot}$ of iron \citep[e.g.][and reference
therein]{gibson97b}. The explosion energy of each SN of both types is
assumed to be $E_{\rm SN}=10^{51}$ erg. See \citet{marcolini2007} for
more details of the chemical evolution implementation.

\begin{table*} 
\centering   
\begin{minipage}{140mm}   
\caption{Galaxy parameters} 
\label{tab:galaxy}  
\begin{tabular} {|c|c|c|c|c|c|c|c|c|}   
\hline   
         $M_{\rm h}  $ &     
         $r_{\rm h,c}$ & $r_{\rm h,t} $ &   
         $\rho_{\rm h,0}$ &  $M_{\star}$  &
         $r_{\star, \rm c} $ &  $r_{\star, \rm t}$ & 
         $\rho_{\star, \rm 0}$  & $(M/L_{\rm V})_0^a$ \\
      
         ($10^6$ M$_{\odot}$) &  (pc)  & (pc) &  ($10^{-24}$ g cm$^{-3}$) 
         & ($10^5$ M$_{\odot}$) & (pc)  & (pc) & ($10^{-24}$ g cm$^{-3}$) \\
  
\hline   

         62  & 300 & 1222 & 4.3 & 5.6 & 130 & 650  & 1.0 & 80 \\  

\hline 

\end{tabular}   
\par  
\medskip 
 $^a$ Value calculated inside $r_{\star, \rm t}$.
\end{minipage}   
\end{table*}   
  
\subsection{dSphs evolution}

We found that, although the total energy released by the SNe~II
explosions is larger than the binding energy of the ISM, efficient
radiative losses enable the galaxy to retain most of its gas, which
thus remains available for the aforementioned prolonged SFH.

The burst of SNe II associated with each stellar burst pushes the bulk
of the ISM to the outskirt of the galaxy. Once the explosions cease
($\sim$30~Myr after each star burst episode), the ISM flows back
towards the centre of the galaxy; when the next burst occurs, the gas
is pushed outwards again.  This oscillatory behaviour leads to a
rather efficient and homogeneous pollution of the ISM by the SNe~II
ejecta. We note that as the SFH has been fixed {\it a priori}, it has
no direct relation to the gas reservoir within the galaxy. However,
stars do form during the quiescent periods between bursts, when the
gas has ``settled", and an {\it a posteriori} consistency for the SFH
is recovered.  As the galaxy in the present model does not expel the
bulk of its ISM, an external cause, such as the interaction with the
MW \citep{mayer2006}, must be invoked at some point during the
evolution to deprive the galaxy of its ISM (as the observed Local
Group dSphs are essentially devoid of gas).

\begin{figure*}    
\psfig{figure=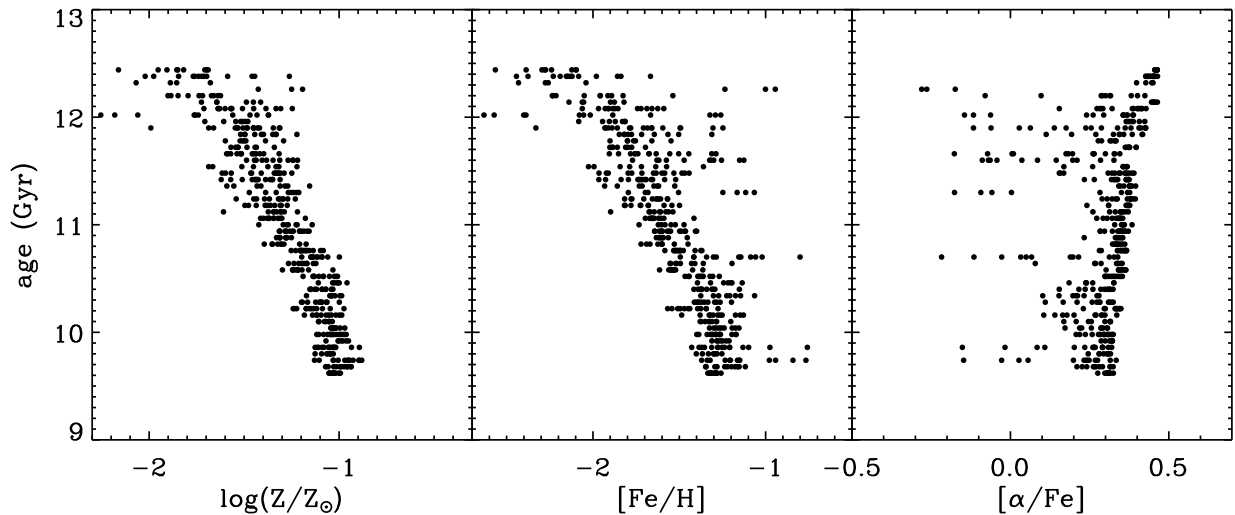,width=0.95\textwidth}    
\caption{Age-$Z$, age-[Fe/H] and age-[$\alpha$/Fe] distributions for
500 sampled stars within one core radius of the simulated volume. Note
the metallicity and [$\alpha$/Fe] spread amongst coeval stars.}
\label{fig:zstelleage} 
\end{figure*}

Given their lower rate, SNe~Ia do not significantly affect the general
hydrodynamical behaviour of the ISM, but their role is relevant for
the chemical evolution of the stars. Because of their longer
evolutionary timescales, SNe~Ia progenitors created in previous
starbursts continue to explode during the subsequent quiescent
periods, when the gas is flowing back into the central region. During
these periods the higher ambient gas density (together with the lower
SNe~Ia explosion rate) cause the SNe~Ia remnants to be isolated from
one another, forming chemically inhomogeneous pockets (hereafter, we
refer to these remnants as ``SNe Ia pockets'').  These pockets are
``washed out" by successive phases of expansion and collapse of the
ISM, due to the effects of SNe~II, but new pockets may form during the
quiescent phases between consecutive starbursts. At odd with the
ejecta of SNe~II, SN~Ia debris is rich in iron and deficient in
$\alpha$-elements. Thus, stars forming in the SN~Ia pockets possess
lower [$\alpha$/Fe] ratios and higher [Fe/H] ratios than those formed
elsewhere. This effect is particularly important for the chemical
evolution of the central galactic region, where the SN~Ia rate is
greater, and higher ambient gas density is achieved.

In the next section we highlight the differences between stars located
in the centre and stars located in the outskirts of the galaxy,
testing our results against several observational constraints.

\section{results}
\label{sec:res}

Although our model was tailored to the Draco dwarf galaxy, it provides
a valuable baseline with which to illustrate several generic
properties of dSphs.

Figure~\ref{fig:zstelle} shows the iron distribution function (IDF) of
the long-lived stars (with masses M $<0.9$ M$_{\odot}$) at three
different times (1, 2 and 3 Gyrs) for the whole dSph (solid line) and
within two (dashed line) and one (dotted line) core radii. Initially,
when the ISM is rather metal-poor, stochastic effects dominate the
chemical evolution owing to the still low number of SNs explosions and the
inhomogeneous dispersal of their ejecta; as a consequence, the IDF is
rather broad and symmetric, as shown in the first panel of the
figure. With time, more and more metals pollute the ISM and are
homogeneously dispersed by the cyclic behaviour of the gas flow
described in the previous section; the newly forming stars are more
iron-rich, and the IDF peak moves toward the high-metallicity side of
the distribution.

Despite the absence of winds, only $\sim 20\%$ of
the SNe ejecta is present within the star forming region.
The remainder of the metals, although still gravitationally bound to the
galaxy, is driven outwards by the effects of SNe explosions
(mimicking, to some degree, an outflow)
and does not contaminate the subsequently forming stars.

Even with the homogenising action of SNe~II, a degree of inhomogeneous
pollution persists due to the newly created SNe~Ia pockets; the
iron-rich stars forming in such pockets populate the ``tail''
at high [Fe/H] in the IDF. As apparent from
Fig.~\ref{fig:zstelle}, this tail is primarily found within
two core radii of the galactic centre, 
and establishes a chemical difference between the
central and outlying stellar populations.

\begin{figure}    
\begin{center}    
\psfig{figure=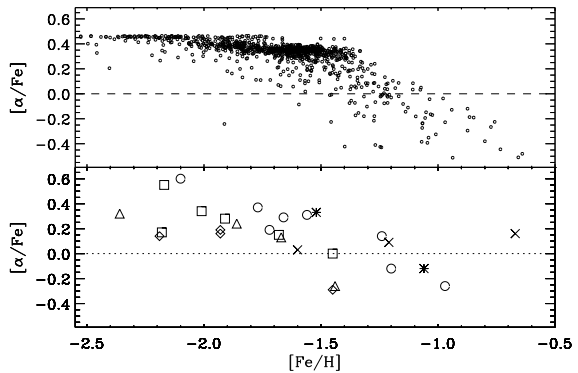,width=0.45\textwidth}    
\end{center}   
\caption{Abundance ratio [$\alpha$/Fe] versus [Fe/H] for 500 sampled
stars of the reference model at $t=2$ Gyr, compared to a dataset of 28
dSph stars (Draco: triangles ; Fornax: crosses; Leo I: asterisks;
Sextans: diamonds; Sculptor: circles; Ursa Minor: squares) collected
from the literature (see text for details). The stars are sampled
as in Fig. \ref{fig:zstelleoss}.}
\label{fig:zstelleosscon} 
\end{figure}

As noted in Section \ref{sec:genpi}, the stars forming in the SNe~Ia
pockets are characterised, in addition to their higher iron abundance,
by their low values of [$\alpha$/Fe] (down to $\sim -0.5$). This is
apparent in Fig. \ref{fig:zstelleoss} which shows the distribution in
the [$\alpha$/Fe]-[Fe/H] plane of a sample of 500 stars representative
of the whole stellar population (upper panels) and of the stars within
one core radius (lower panels).\footnote{One of the outputs of our
code at any time is the three-dimensional stellar distribution
$\psi(f,o,r)$ (where, for each star, $f\equiv$ [Fe/H], $o\equiv$
[O/Fe], and $r$ is the position). We can obtain $\Phi(f)=\int dr\int
\psi(f,o,r)do$ (shown in Fig. \ref{fig:zstelle}) and $\phi(f,o)=\int
\psi(f,o,r)dr$. The random sampling shown in Fig.~\ref{fig:zstelleoss}
is done as follows: we first extract randomly a value of [Fe/H] from
the distribution $\psi$, then a value of [$\alpha$/Fe] is randomly
extracted from the $\phi$ distribution calculated for the fixed value
of [Fe/H]. Similar procedures have been followed to obtain the
samplings shown in
Fig.~\ref{fig:zstelleage},~\ref{fig:zstelleosscon},~\ref{fig:zraggio}
and \ref{fig:alpharaggio}.} As expected, a higher fraction of
$\alpha$-depleted stars is present in the central region, where SNe~Ia
pockets are preferentially found.  Indeed, the fractions of stars with
[$\alpha$/Fe]$<$0.1 present after 1~Gyr within one core radius, two
core radii, and the entire galaxy are $\sim$20$\%$, $\sim$15$\%$ and
$\sim$10$\%$, respectively.  These values decrease with time, and at
$t=3$ Gyr they are reduced by a factor $\sim$30\%. This reduction is
due to the decreasing relevance of the contribution of the single
pockets, as the mean value of [Fe/H] increases with time while the
iron content of the pockets is always the same. In fact, from
Fig. \ref{fig:zstelleoss} (as well as from Fig. \ref{fig:zstelle}) it
is clear that the maximum value of [Fe/H] never exceeds $\sim
-0.7$. Actually, this value is attained by the stars forming
within each pocket, where the amount of iron delivered by the SN~Ia is
constant (0.74 M$_{\odot}$), and the amount of H does not vary
appreciably because the stars form during the quiescent periods, when
the ISM settled within the galactic core recovers substantially to the
same densities. As a consequence, the maximum obtainable value of
[Fe/H] is rather insensitive to the ISM mean metallicity, and depends
instead on the details of the dSph model.  In the particular model we
are discussing, the SNe~Ia inhomogeneous pollution is expected to
become irrelevant at a nearly solar value of [Fe/H]. The effects of
the inhomogeneity are thus expected to be crucial in the metal-poor
dwarf galaxies \citep{marcolini2007}, but not in those more
metal-rich.

The stars in Fig. \ref{fig:zstelleoss} are arranged in two branches, a
nearly horizontal plateau and an oblique ``ramp'' connected to the
plateau by a ``knee''. The stars located in the upper envelope of the
plateau are mostly polluted by the ejecta of SNe~II, while the stars
below the plateau, as well those on the ramp, are also inhomogeneously
(but significantly) polluted in different measure by SNe Ia \citep
[see][for more details]{marcolini2006}.  As the mean iron content of
the ISM increases with time, the knee in the [$\alpha$/Fe]-[Fe/H]
diagram moves rightward (toward larger values of [Fe/H]) and, to a
lesser extent, downward (to lower [$\alpha$/Fe]).

A further evidence of the inhomogeneous pollution is given in
Fig.~\ref{fig:zstelleage}, where we plot the age-$\log(Z/Z_{\odot})$,
age-[Fe/H] and age-[$\alpha$/Fe] relations for a randomly-drawn sample
of 500 stars formed within one core radius. In the first two panels a
striking spread in the metallicity $Z$ and in the iron content of the
star is apparent, which tends to reduce with time, as expected.  The
[Fe/H] spread is more accentuated because stars formed inside SNe~Ia
pockets have a much higher iron content, but essentially the same $Z$,
than the stars formed elsewhere, since SNe~Ia deliver a significant
amount of iron but a negligible fraction of metals in comparison to
SNe~II (for example, after 3~Gyr the SNe~Ia ejecta accounts for 30\%
of the iron produced by all stars, but only for 2\% of the metals).

\begin{figure*}    
\psfig{figure=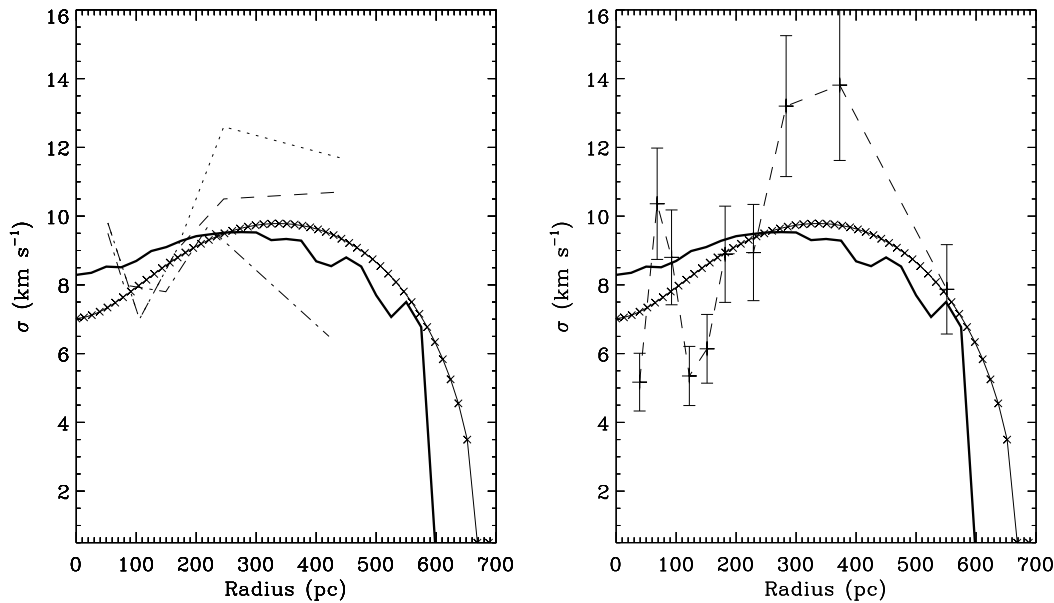,width=0.90\textwidth}    
\caption{The radial profile of the 3D velocity dispersion (solid line
with crosses) and projected velocity dispersion using 1000 sampled
stars of our model (thick solid line) obtained solving the
Jeans-equation compared to the velocity dispersion of Draco dSph
obtained by \citet[][left panel]{lokas2005} for three samples of 207
(dotted), 203 (dashed) and 189 (dash-dotted) stars, and compared to the
velocity dispersion obtained by \citet[][right panel]{munoz2005}.}
\label{fig:sigma} 
\end{figure*}

It is commonplace to assume that stars have chemical elements in solar
proportion, thus leading to an implicit equivalence between the IDF
and the metal distribution function (MDF).  That said, a comparison
between the first two panels of Fig.~\ref{fig:zstelleage} indicates
that the two distributions are different \citep[see][for more details
about this point and how it influences the color-magnitude
diagrams]{marcolini2007}.

Finally, the third panel of Fig.~\ref{fig:zstelleage} illustrates the
evolution of [$\alpha$/Fe] and shows that the SNe~Ia pockets start to
form quite early in the evolution ($\sim$300 Myr after the first star
formation episode, as the SNe~Ia rate starts to become appreciable)
and the tendency of its spread to decrease with time (see also the
lower panels of Fig. \ref{fig:zstelleoss}).

\section{Comparison with Data}
\label{sec:data}
 
In Fig.~\ref{fig:zstelleosscon}, we compare the chemical
characteristics of the stars of the model (at 2 Gyr) with data
published by several authors \citep{shetrone2001, shetrone2003,
tolstoy2003, geisler2005} for different dSphs (Draco, Fornax, Leo I,
Sextans, Sculptor and Ursa Minor) observed using VLT/UVES.

Although the dataset is a collection of values of different dSphs, the
general pattern of the observed [$\alpha$/Fe]-[Fe/H] diagram is
reproduced quite well by our single model. Due to the small number
statistics in published results it is difficult to draw definitive
conclusions. \citet{tolstoy2003} suggested that the similar abundance
patterns across several dSph galaxies was evidence for very similar
chemical evolution histories, similar initial conditions and similar
initial mass functions, despite their different SFHs, although there
is also evidence for differences \citep{venn2004, tolstoy2006,
latarte2007}. Larger samples coming from FLAMES high resolution data
sets should make this issue clearer (Hill et al. in prep; Venn et
al. in prep), greatly increasing the statistics inside each single
galaxy, including the Sculptor and Fornax dSphs. Indeed, one should
compare the model with a dataset for {\it each} dwarf galaxy as the
corresponding "knee" can start at different [Fe/H] values depending
upon the chemical enrichment history of the dSph
citep{tolstoy2006,latarte2007}. However, for the aim of this paper,
the literature data are sufficient to highlight the main
characteristics (note that the future larger samples will overlay the
present UVES sample). Here, we simply wish to highlight the point that
$\alpha$-depleted stars are observed in several dSphs, just as
predicted by the model.

\begin{figure*} 
\begin{center}
\psfig{figure=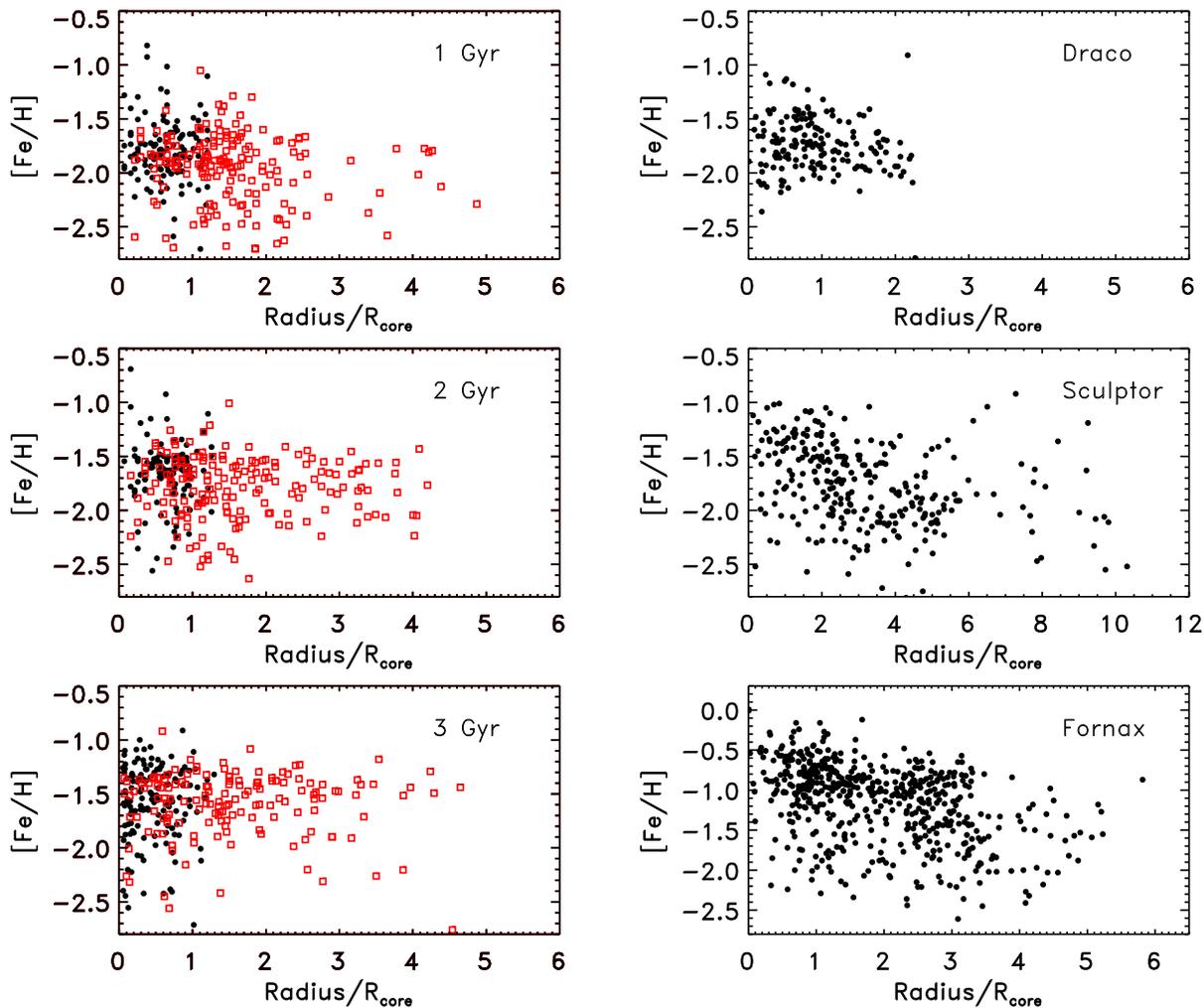,width=0.95\textwidth}
\end{center} 
\caption{Left panels: [Fe/H] projected radial distributions of 300
sampled stars at three different times (1, 2 and 3 Gyr). Right panels:
observed [Fe/H] radial distributions for 87 stars in Draco
\citep{faria2007}, 308 stars in Sculptor \citep{tolstoy2004} and 562
stars in Fornax \citep{battaglia2006}.  The empty squares represent
stars with $\sigma>8.5$ km s$^{-1}$ while black dots are stars with
$\sigma<8.5$ km s$^{-1}$.}
\label{fig:zraggio} 
\end{figure*}

Actually, $\alpha$-depleted stars can also be explained within the
framework of the chemical model of \citet{lanfranchi2004,
lanfranchi2007}, in which the [$\alpha$/Fe] decrease at larger values
of [Fe/H] is the consequence of a galactic wind which reduces the
amount of available gas; the formation of new stars is thus tempered,
along with the $\alpha$-elements delivered by SNe~II. SNe~Ia, instead,
having progenitors with longer evolutionary timescales
\citep{matteucci1986a}, continue to inject iron into the ISM, lowering
the [$\alpha$/Fe] ratio of the gas. In this case the $\alpha$-depleted
stars should be the youngest ones, with the largest values of [Fe/H],
and exhibiting a small spread in age. Preliminary results of
[$\alpha$/Fe] measurements of $\sim 90$ stars in Sculptor and $\sim
55$ stars in Fornax recently published by \citet{tolstoy2006} seem to
indicate that $\alpha$-depleted stars are found at any [Fe/H]$> -1.8$
down to [Fe/H]$\sim-0.6$ inside each galaxy, in good agreement with
our model.

In Fig.~\ref{fig:sigma} the projected radial velocity dispersion
profiles of the model are plotted and tested against data inferred for
Draco by \citet{munoz2005} and \citet{lokas2005}. The velocity
dispersion profile is calculated solving the Jeans equation and
assuming velocity isotropy ($\beta=0$). Different choices of the
$\beta$ parameter can change the results dramatically, even if there
is no valid motivation, {\it a priori}, for choosing highly tangential
or radial orbits; as such, $\beta=0$ appears a fair conservative
value. For example \citet{mateo2007} found that for Leo~I stars, those
situated more towards the centre appear to have isotropic kinematics,
while only those located outside a ``break radius'' of about 500~pc
exhibit significant velocity anisotropy. These authors also suggest
that the break radius represents the location of the tidal radius of
Leo~I.

In the same figure, we also plot the actual 3D velocity dispersion
profile within the galaxy (solid line with crosses).  While the
projected profile is essentially flat (within 1~km~s$^{-1}$) up to
500~pc, the the actual profile shows a central minimum, $\sim$ 3km
s$^{-1}$ below the maximum value attained in the outer regions.  The
tendency of the central stars of the model to have somewhat lower
values of the velocity dispersion is emphasised in the left panels of
Fig.~\ref{fig:zraggio}, where we show the projected radial positions
and the ratios [Fe/H] of a sample of 300 stars: here the empty squares
represent stars with $\sigma>8.5$ km s$^{-1}$ while black dots are
stars with $\sigma<8.5$ km s$^{-1}$. These diagrams also show that
[Fe/H] rich stars are found preferentially in the central region of
the system (and only with lower velocity dispersion), while the main
Fe-poor population is more smoothly distributed. This result can be
easily understood: while the iron-rich stars form preferentially in
the SNe Ia pockets which are more concentrated toward the center, the
stars with lower values of [Fe/H] form throughout the galaxy from an
ISM whose pollution does not depend strongly on the radius because of
the dynamically-driven homogenisation (cf. Section \ref{sec:genpi}).
As a consequence, two different stellar populations are found in our
model, with iron-poor stars showing higher velocity dispersion and
iron-rich (centrally concentrated) stars showing lower velocity
dispersion. A similar anti-correlation between the iron content and
the velocity dispersion has been actually observed in Sculptor
\citep{tolstoy2004} and Fornax \citep{battaglia2006} dSphs (and found
in preliminary models that we adapted to these galaxies).

From the left panels in Fig.~\ref{fig:zraggio} it is apparent that the
metallicity gradients occurring in our model are rather small, at
variance with monolithic dissipative models of elliptical galaxy
formation. In these models strong gradients form as a consequence of
existing radial flows carrying inward the heavy elements produced in
the outer regions \citep{lars74}. In our model, instead, because of
the small size of dSphs, the gas flows powered by the SNe II tend to
homogenize the ISM, thus inhibiting the formation of chemical
gradients (cf. section \ref{sec:genpi}); the present mild [Fe/H]
gradient is mainly due to SNe Ia, as discussed in \ref{sec:res}.
Indeed, being that the oxygen (and $\alpha$-elements in general)
is produced for the most part by SNe II, the [O/H] gradient is even
shallower than that of [Fe/H] \citep{marcolini2006}.

From the left panels in Fig.~\ref{fig:zraggio} it is also apparent
that the small metallicity gradient (0.2-0.4 dex; see Marcolini et
al. 2006) tends to decrease with time as the amount of iron in the ISM
increases and the stars forming at larger radii acquire increasing
values of [Fe/H].  This tendency seems to be confirmed by the stellar
radial distributions of Draco \citep{faria2007}, Sculptor
\citep{tolstoy2004} and Fornax \citep{battaglia2006} (right panels in
Fig.~\ref{fig:zraggio}): these dSphs seem to possess shallower
gradients, as their SFH \citep[e.g.][ and reference
therein]{buonanno1999, dolphin2002, tolstoy2003} is more
prolonged. While this behaviour can be described only qualitatively by
the reference model of \citet{marcolini2006}, further models tailored
to Sculptor and Fornax are planned, in order provide a more
appropriate and direct comparison to the data.  Nevertheless, our
model seems to be in good agreement with the ``central concentration''
of the ``small [Fe/H] tail'' found by \citet{faria2007} for Draco.

\begin{figure}    
\begin{center} 
\psfig{figure=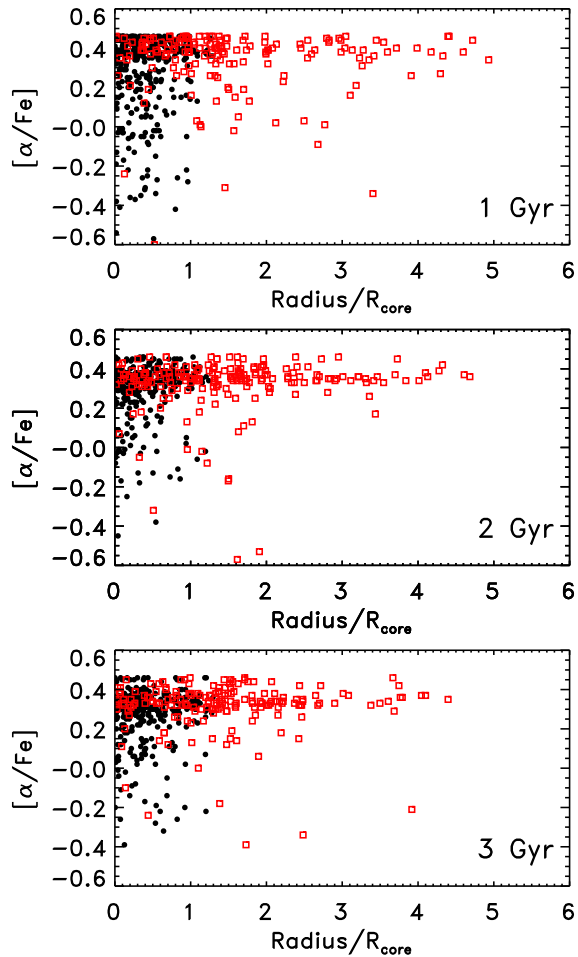,width=0.40\textwidth}
\end{center} 
\caption{[$\alpha$/Fe] projected radial distributions of 500 sampled
stars at three different times (1, 2 and 3 Gyr).  The empty squares
represent stars with $\sigma>8.5$ km s$^{-1}$ while black dots are
stars with $\sigma<8.5$ km s$^{-1}$.}
\label{fig:alpharaggio} 
\end{figure}

We stress that the presence in the galactic center of an iron-rich
stellar population could be explained in principle also by other
models. In the framework depicted by \citet{mori2002} the SNe feedback
following the initial star formation blows out the gas and stops star
formation temporarily. After SNe feedback becomes weaker, more
metal-rich gas can return and form a new generation of stars
distributed closer to the center. As pointed out in Section
\ref{sec:introduction}, however, the model of \citet{mori2002} is not
suitable in the case of the prolonged SFHs often encountered in
dSphs. A more metal rich central population of stars is also obtained in
the model of \citet{kawata2006} as the result of a steep metallicity
gradient within a single population, induced by dissipative collapse
of the gas component. This is the same mechanism that is suggested to
explain the metallicity gradient for larger spheroidals, i.e.  normal
elliptical galaxies. It is interesting to note that in these galaxies
the observed gradient of [$\alpha$/Fe] has a mean value close to zero,
and that it does not correlate with galactic properties
\citep[e.g.][]{sanchez2007,pipino2007}. Thus, within the scheme
proposed by \citet{kawata2006}, dSphs are not expected to present a
well defined [$\alpha$/Fe] gradient. On the contrary, in the model of
\citet{marcolini2006} $\alpha$-depleted stars are centrally segregated
(see Fig.~\ref{fig:alpharaggio}) as they are correlated with the
Fe-rich stars which are preferentially found in the central region.
Thus the observations of the [$\alpha$/Fe] radial profile in dSphs can
potentially select between the model of \citet{marcolini2006} and the
scheme proposed by \citet{kawata2006}. As an aside, the model of
\citet{marcolini2006} also predicts that the most $\alpha$-depleted
stars have lower velocity dispersions, as illustrated in
Fig.~\ref{fig:alpharaggio}.

We point out that, although the proximity of the MW may change the
stellar orbits (expecially in the outskirts) of a local dSph, a
segregation of the $\alpha$-depleted stars should still be present
because such stars are located within two core radii, a region only
slightly affected by the tidal interaction
\citep[e.g.][]{gao2004,penarrubia2008}.

\section{Conclusions}
\label{sec:discussion}

We analysed the outcome of the 3D hydrodynamical model of the Draco
dSph of \citet{marcolini2006}, and found that the results can provide
a useful, if qualitative, description of the larger population of
generic Local Group dSphs. We can summarise our major findings as
follows:

$i$) Evidence is mounting about the effects of inhomogeneous pollution
on the chemical history of the stars in dSphs \citep{koch2008}. The
effect of the inhomogeneous pollution of SNe~Ia found in
\citet{marcolini2006} is more important in the central region (within
two core radii) of dSphs. Here the stars forming in the SNe~Ia pockets
can amount to $\sim 20\%$ of the total.

$ii$) The chemical homogenisation of the ISM by SNe II, together with
the combined inhomogeneous pollution by SNe~Ia, naturally accounts for
a radial segregation of Fe-rich stars (with depleted [$\alpha$/Fe]
ratios) in the central regions of dSphs. Stars with this abundance
pattern have been observed in the center of several dSphs
\citep{shetrone2001, shetrone2003, bonifacio2004, monaco2005,
tolstoy2006, koch2008}. As discussed, the most Fe-rich stars in our
model are also $\alpha$-depleted, and thus the same radial segregation
should be observed to test and prove our model.

$iii$) At odd with the projected stellar velocity dispersion radial
profile, the actual one shows a depression in the central region,
where the iron-rich stars are partly segregated. This naturally
entails two different stellar populations with an anti-correlation
between [Fe/H] and velocity dispersion, which has been 
observed in the case of the Sculptor dSph \citep{tolstoy2004} and the
Fornax dSph \citep{battaglia2006}.

In a forthcoming paper we will compare new simulations tailored to
specific dSphs (e.g. Fornax and Sculptor) with the properties of
hundreds of stars observed with VLT-FLAMES that will be available soon
\citep[][and references therein]{venn2005, tolstoy2006}. The new
models will take into account the yields from low- and
intermediate-mass asymptotic giant branch stars, as well as the
interaction with our Galaxy, in order to describe how the dSphs lose
their gas. For example, an interaction with the MW seems to be
particularly important in the evolution of the Sagittarius dwarf.
Preliminary results show that the iron-rich, $\alpha$-depleted stars
in this galaxy are consistent with a SF strongly affected by this
interaction, as well as with the formation {\it in situ} of M54 as the
core of this galaxy.

\section*{Acknowledgments}

We thank the anonymous referee for his/her useful comments that
improved the presentation of the paper. We also kindly thank Eline
Tolstoy for reading the manuscript and for useful discussion, we are
also in debt with Ricardo {Mu{\~n}oz} for giving us the data to make
Fig.~\ref{fig:sigma}. This research was undertaken as part of the
Commonwealth Cosmology Initiative (CCI:www.thecci.org).  AD
acknowledges financial support from National Institute for
Astrophysics (INAF). The simulations were run at the CINECA
Supercomputing Centre, and supported through the award of grant from
INAF-CINECA.  Analysis has been undertaken at the UCLan HPC Facility.
    

\bibliographystyle{mn2e}    
\bibliography{dsphe_refs}

\label{lastpage}    
\end{document}